\documentclass[aip,reprint,jcp]{revtex4-2}

\usepackage[version=3]{mhchem} 
\usepackage[%
  colorlinks=true,
  urlcolor=blue,
  linkcolor=blue,
  citecolor=black
]{hyperref}
\usepackage{graphicx}
\usepackage{dcolumn}
\usepackage[normalem]{ulem}
\usepackage{bm}
\usepackage[utf8]{inputenc}
\usepackage[T1]{fontenc}
\usepackage{mathptmx}
\usepackage{amsmath}
\renewcommand{\selectlanguage}[1]{}
\DeclareUnicodeCharacter{2009}{\,}
\DeclareUnicodeCharacter{2032}{'}

\usepackage{xcolor}

\makeatletter
\def\@email#1#2{%
 \endgroup
 \patchcmd{\titleblock@produce}
  {\frontmatter@RRAPformat}
  {\frontmatter@RRAPformat{\produce@RRAP{*#1\href{mailto:#2}{#2}}}\frontmatter@RRAPformat}
  {}{}
}%
\makeatother

\begin{document}

\author{Pierre J. Walker}
\altaffiliation{These authors contributed equally.}
\affiliation{Division of Chemistry and Chemical Engineering, California Institute of Technology, Pasadena, CA 91125, United States}

\author{Ananya Venkatachalam}
\altaffiliation{These authors contributed equally.}
\affiliation{Department of Chemistry, Harvey Mudd College, Claremont, CA 91711, United States}

\author{Samuel Varner}
\altaffiliation{These authors contributed equally.}
\affiliation{Division of Chemistry and Chemical Engineering, California Institute of Technology, Pasadena, CA 91125, United States}

\author{Bilin Zhuang}
\affiliation{Department of Chemistry, Harvey Mudd College, Claremont, CA 91711, United States}
\email{bzhuang@g.hmc.edu}

\author{Zhen-Gang Wang}
\affiliation{Division of Chemistry and Chemical Engineering, California Institute of Technology, Pasadena, CA 91125, United States}
\email{zgw@caltech.edu}

\title{Stockmayer Fluid with a Shifted Dipole: Bulk Behavior}

\begin{abstract}
Displacing the point dipole from the center of a Stockmayer particle is a simple geometric modification that has been examined previously, yet its physical consequences for liquid structure, dielectric response, and phase behavior remain only partially understood. Here, we combine molecular dynamics simulations with analytical theory to provide a unified mechanistic interpretation of how dipole displacement reshapes microscopic correlations and propagates to macroscopic thermodynamic properties. We show that shifting the dipole breaks the fore--aft symmetry of the local dipolar field, producing only minor changes in radial packing but pronounced alterations in angular structure within the first solvation shell. Enhanced accumulation and alignment near the head of the particle are accompanied by frustrated orientational correlations near the tail, leading to broadened angular distributions and shift of preferred configurations away from axial positions at strong dipole coupling. These structural asymmetries weaken cooperative dipolar ordering and cause a systematic reduction in the dielectric constant, despite locally stronger interactions. For sufficiently large shifts, the dielectric response approaches the Debye mean--field limit, indicating that the effect of dipole correlations effectively suppressed. The same geometric frustration governs vapor--liquid equilibria. While increasing dipole strength raises the critical temperature as in the conventional Stockmayer fluid, introducing even modest shifts disrupts the highly ordered polarized liquid states that emerge at strong coupling at high dipole strengths, going as far as to suppress ferroelectric--like ordering at high dipole strengths. Predictions from a reparameterized Co-Oriented Fluid Functional Equation for Electrostatics (COFFEE) theory capture these trends within its domain of applicability and highlight the direct connection between local orientational structure and macroscopic observables. Overall, this work demonstrates that the location of the electric dipole in the molecule—not merely their magnitude—acts as a powerful control parameter in dipolar liquids, providing a clear physical framework for interpreting and exploiting geometric frustration in electrostatic fluids.
\end{abstract}

\maketitle

\section{Introduction}


Electrostatic interactions govern liquid structure and thermodynamic behavior across scales, from local solvation structures to macroscopic response functions, for example, the dielectric constant.\cite{zhuang_like_2021} The cross-scale connection is vital for the interpretation and modeling of polar liquids and mixtures. This coupling between scales is especially important when one considers both the bulk behavior of solvents, such as the preferential solvation of ions\cite{nakamuraIonSolvationLiquid2012,gregory_understanding_2022}, and at liquid interfaces, where orientational ordering can generate local electric fields that bias reaction pathways,\cite{jubbEnvironmentalChemistryVapor2012, tobiasSimulationTheoryIons2013, ruiz-lopezMolecularReactionsAqueous2020, tangMolecularStructureModeling2020, martins-costa_electrostatics_2023, martins-costa_effect_2025, liang_water_2023, xie_harnessing_2024, kang_theoretical_2025, shiWaterStructureElectric2025}. The present study focuses on equilibrium bulk structure and dielectric response, where connections between microscopic correlations and observables can be most clearly established.

The magnitude and physical origin of anomalously strong interfacial fields and their accompanying bulk properties remain actively debated, emphasizing the need for minimal, controllable models that enable rigorous study of electrostatics, packing, and orientational correlations .\cite{kol_long-range_2025, shirley_reevaluating_2025} Dipolar model fluids provide a convenient system to isolate the effects of excluded-volume packing and orientational interactions. Among such models, the Stockmayer fluid is an especially effective minimal model of a polar liquid .\cite{stockmayerSecondVirialCoefficients1941,shock_solvation_2020,liu_minimal_2024} Simulations and existing theory have used the Stockmayer model to connect microscopic pair correlations to collective behavior, including in regimes that exhibit strong orientational ordering at large dipole strengths .\cite{pollock_static_1980, adams_static_1981, groh_ferroelectric_1994, bartke_dielectric_2006} Stockmayer-type models have also been used to investigate liquid--vapor interfaces and the impact of external fields on surface polarization .\cite{eggebrecht_liquidvapor_1987, paul_liquidvapor_2003, samin_vapor-liquid_2013, moore_liquid-vapor_2015} Interactions in this model are simple enough to interpret mechanistically and yet are able to capture nontrivial coupling between local structure and macroscopic observables. 

A primary limitation of the Stockmayer fluid model is its treatment of molecules as perfect spheres with centrally located point dipoles. While mathematically convenient, this idealized symmetry fails to describe the inherent asymmetry of most polar molecules. In reality, few molecules exhibit such fore–aft symmetry; instead, they often possess off-center charge distributions where the electrostatic and geometric centers do not coincide. To bridge this gap, the shifted Stockmayer fluid (sSF) model introduces a minimal source of electrostatic asymmetry by displacing the dipole away from the Lennard-Jones center, providing a more physically grounded representation of molecular geometry. This model was first introduce by Kusaka et al.\cite{kusaka_ion-induced_1995} to examine the sign effect in ion-induced nucleation of water-droplet in the atmosphere. Kantorovich and co-workers\cite{kantorovichFerrofluidsShiftedDipoles2011,kantorovichMagneticParticlesShifted2011,klinkigtClusterFormationSystems2013,weeberMicrostructureMagneticProperties2013} used a similar model fluid to examine cluster formation in ferrofluids, focusing on large shifts and high dipole strengths. Particularly relevant to this study is the work by Langenbach and co-workers\cite{langenbach_simultaneous_2015,langenbach_co-oriented_2017,kohns_relative_2020,kohns_critical_2021,marx_phase_2022,marx_vapor-liquid_2023} who developed the Co-Oriented Fluid Functional Equation for Electrostatic Interactions (COFFEE) framework to model the fluid-phase behavior of the shifted Stockmayer fluid, using information from the orientational distribution functions to determine thermodynamic properties. For small dipolar shifts and strengths, this work has been used to study the dielectric properties\cite{kohns_relative_2020,kohns_critical_2021} and phase behavior,\cite{langenbach_co-oriented_2017} even extending the work to mixtures,\cite{marx_phase_2022,marx_vapor-liquid_2023} for this simple model fluid. Collectively, these studies establish that even moderate dipole shifts can substantially impact dielectric response and vapor--liquid equilibria of a dipolar fluid. Despite ongoing work, a fundamental gap remains in the physical intuition underlying the effect of introducing a dipole shift and the resulting macroscopic changes. Much of the existing literature emphasizes reproducing and predicting simulation results, while the underlying driving forces remain unresolved. For example, it is not clear whether increasing the magnitude or shift of the dipole increases global order or frustrates it, or how local structural asymmetries translate into the dielectric response. 

We build on prior studies of sSF but shift the emphasis to developing an understanding of the bulk physics of this system. To this end, we leverage molecular dynamics simulations, as discussed in Section \ref{sec:md}. We also evaluate the suitability of the COFFEE theory for studying the bulk behavior of sSF systems, as discussed in Section \ref{sec:coffee}. We then examine the structural properties of the sSF across a wide range of dipole shifts and strengths, developing intuition for the changes in the orientational-radial distribution functions in Section \ref{sec:struct}. Using this intuition, we examine changes in the dielectric response and phase behavior of sSF fluids in Sections \ref{sec:dielectric} and \ref{sec:vle}, respectively. We also note that this is a companion piece to a following study examining the vapor--liquid interfacial behavior in sSF systems.\cite{varnerStockmayerFluidShifted2025}


\begin{figure}[ht!]
    \centering
        \includegraphics[width=0.4\linewidth]{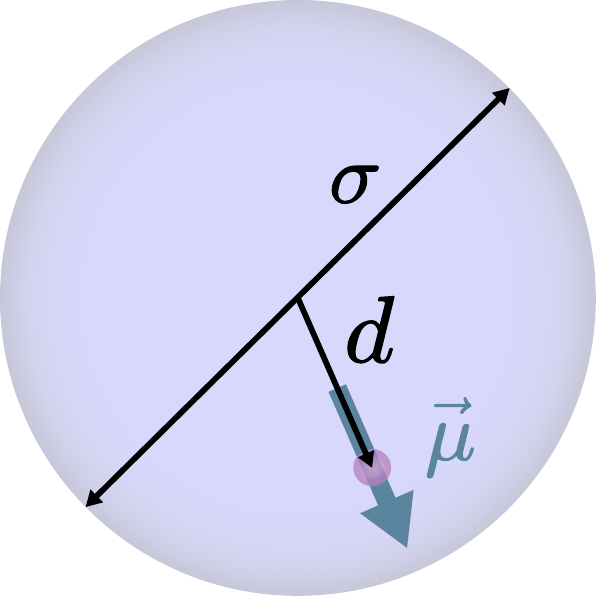}    
    \caption{Model of a shifted Stockmayer fluid (sSF). Compared to the regular Stockmayer fluid, the particle has a dipole $\vec{\mu}$ shifted from the center by a distance $d$ in the direction of the dipole (depicted as the blue arrow). The excluded-volume interaction between the particles is described by a regular Lennard-Jones potential with characteristic length scale $\sigma$, and the electrostatic interaction is the dipole-dipole interaction between the shifted dipoles.}  
    \label{fig:shifted-stockmayer}
\end{figure}

\begin{figure*}[ht!]
    \centering
        \includegraphics[width=1\linewidth]{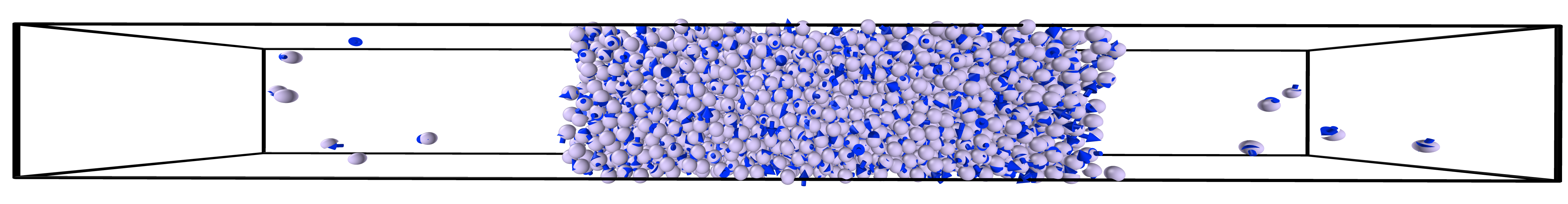}
    
    \caption{Slab system of shifted-dipole Stockmayer particles. The Lennard-Jones beads are visualized in light purple and the shifted dipoles in dark blue. Ghost particle have been hidden.}  
    \label{fig:interfacialbox}
\end{figure*}


\section{Methodology}

\subsection{System}

We consider a system of $N$ particles at constant volume $V$ and temperature $T$. 
An example particle $i$ is illustrated schematically in Fig.~\ref{fig:shifted-stockmayer}, whose geometrical center is located at $\mathbf{r}_i$ and whose dipole moment $\pmb{\mu}_i$ is shifted by a distance $d$ from this center along the dipole vector. That is, the dipole of the particle is located at $\mathbf{r}^{d}_i = \mathbf{r}_i +  \hat{\pmb{\mu}}_i d$, where $\hat{\pmb{\mu}}_i$ is a unit vector in the direction of $\pmb{\mu}_i$. The particles interact through a truncated Lennard--Jones (LJ) interaction between the geometrical centers and a long--range dipole--dipole interaction between the displaced point dipoles. The LJ interaction is given by:
\begin{equation}
    \label{eq:lj}
    \phi_{\text{LJ}}(r_{ij}) = 
    \begin{cases} 
    4\epsilon \left[ \left( \frac{\sigma}{r_{ij}} \right)^{12} - \left( \frac{\sigma}{r_{ij}} \right)^6 \right], & r_{ij} \leq r_c,\\[6pt]
    0, & r_{ij} > r_c,
    \end{cases}
\end{equation}
where $\epsilon$ is the pair interaction strength, $\sigma$ is the characteristic particle diameter, and $r_{ij} = |\mathbf{r}_j - \mathbf{r}_i|$ is the distance between the geometrical centers of particles $i$ and $j$. The LJ cut-off distance is set to $r_c = 2.5\sigma$, which is well beyond the potential minimum at $r_{\text{min}} = 2^{1/6}\sigma$. 

The electrostatic interactions between the permanent point dipoles are described by the dipole--dipole potential:
\begin{equation}\label{dip}
    \phi_\text{dipole}(\pmb{r}_{ij}^d) = \frac{1}{4\pi\epsilon_0}\left[ \frac{\pmb{\mu}_i\cdot\pmb{\mu}_j}{(r_{ij}^d)^3}-3\frac{\left(\pmb{\mu}_i\cdot\pmb{r}_{ij}^d\right)\left(\pmb{\mu}_j\cdot\pmb{r}_{ij}^d\right)}{(r_{ij}^d)^5}\right]\,,
\end{equation}
where $\epsilon_0$ is the vacuum permittivity, $\pmb{r}_{ij}^d = \mathbf{r}_j^d - \mathbf{r}_i^d$ is the vector connecting the displaced dipoles and $r_{ij}^d = |\pmb{r}_{ij}^d|$ is its magnitude. Here $\pmb{\mu}_i$ and $\pmb{\mu}_j$ denote the dipole moments of particles $i$ and $j$. This model is nearly equivalent to the original Stockmayer fluid model,\cite{stockmayer_second_1941} with the exception of the dipole being shifted off the particle center.

\begin{figure}[ht]
    \centering
    \includegraphics[width=0.9\linewidth]{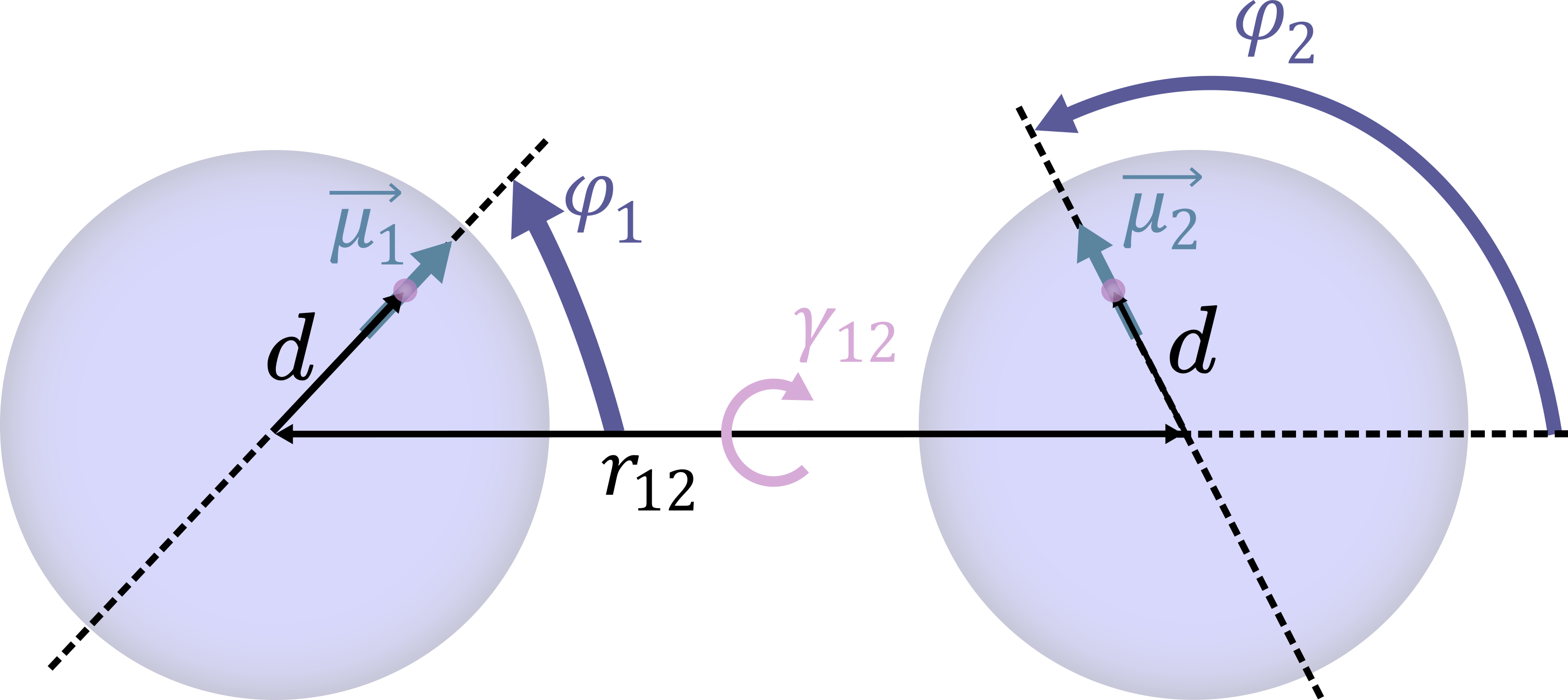}
    \caption{Schematic illustration of angular descriptors for the specific particle pair. The angles $\varphi_1$ and $\varphi_2$ denote the orientation of each dipole moment $\boldsymbol{\mu}_1$ and $\boldsymbol{\mu}_2$ relative to the dipole-displacement vector $\mathbf{r}^{d}$, while $\Theta$ is the angle between the two dipoles. The scalar center-to-center distance is $r$, and $d$ represents the shift of each dipole from the particle center.}
    \label{fig:orientation}
\end{figure}

\subsection{Molecular Dynamics}
\label{sec:md}
\subsubsection{Simulations}

We perform molecular dynamics (MD) simulations of the shifted Stockmayer model using the LAMMPS simulation package ,\cite{thompsonLAMMPSFlexibleSimulation2022} executed on graphics processing units (GPUs) .\cite{brownImplementingMolecularDynamics2011,brownImplementingMolecularDynamics2012} Because no native shifted Stockmayer potential is available in LAMMPS, each molecule $i$ is represented as a rigid dimer composed of two components (Fig.~\ref{fig:shifted-stockmayer}): an LJ particle with characteristic size $\sigma$ and a massless, infinitesimally small ``ghost'' particle carrying a permanent dipole moment $\boldsymbol{\mu}_i$. The two sites are separated by a fixed bond length $d$, such that the dipole is displaced from the geometric center of the particle. The ``ghost'' particle interacts only through dipolar forces, and its dipole is constrained to remain parallel to the rigid bond, pointing radially outward. In the limit $d=0$, this model reduces to the standard Stockmayer fluid. Throughout this work, we vary the dipole magnitude $\mu = |\boldsymbol{\mu}|$, shift $d$, and temperature $T$.

To avoid numerical discontinuities at the LJ cutoff, we apply a constant energy shift $S$ to the potential,
\begin{equation}
    \label{eq:lj_s}
    \phi_{\text{LJ}}^{\text{shifted}}(r_{ij}) = 
    \begin{cases} 
    4\epsilon \left[ \left( \frac{\sigma}{r_{ij}} \right)^{12} - \left( \frac{\sigma}{r_{ij}} \right)^6 \right] + S, & \text{if } r_{ij} \leq r_c\,,\\
    0, & \text{if } r_{ij} > r_c \,,
    \end{cases}
\end{equation}
where $S$ is chosen such that $\phi_{\text{LJ}}^{\text{shifted}}(r_c)=0$. Dipolar interactions are evaluated directly in real space up to a cutoff distance of $8.0\sigma$, beyond which they are computed in Fourier space using the particle--particle particle--mesh (PPPM) method for long--range dipolar interactions .\cite{cerda_p3m_2008} 

To determine vapor--liquid coexistence conditions, we employ slab--geometry simulations with box dimensions $10\sigma \times 10\sigma \times 100\sigma$ containing 3000 physical particles (6000 interaction sites, including the LJ particles and their associated dipolar ``ghosts''), as illustrated in Fig.~\ref{fig:interfacialbox}. Initial configurations are generated by randomly placing particles in the simulation box using Packmol .\cite{martinezPACKMOLPackageBuilding2009} All simulations employ a Nos\'e--Hoover--type thermostat that accounts for both translational and rotational degrees of freedom in rigid bodies,\cite{kamberaj_time_2005,hooverCanonicalDynamicsEquilibrium1985,martynaExplicitReversibleIntegrators1996,martynaNoseHooverChains1992} with a relaxation time $\tau = 100 \Delta t$ and a time step of $\Delta t = 0.005\tau_0$ where $\tau_0=\sqrt{m\sigma^2/\epsilon}$.

To induce slab formation, a weak external force is applied along the $z$-axis to all particles for $5\times10^4$ steps, driving them toward the center of the simulation cell. To suppress interactions between the liquid slab and its periodic images, we apply a slab correction by introducing 12--6 Lennard-Jones walls at the two ends of the simulation box shown in Fig.~\ref{fig:interfacialbox}, together with a vacuum region three times the box length between periodic replicas, thereby rendering the system periodic in two dimensions and nonperiodic in the normal dimension.\cite{yeh_ewald_1999} This slab correction removes spurious interactions between liquid slabs along the $z$-direction while retaining the efficiency of the three-dimensional PPPM solver. After slab formation, the external force is removed and the system is allowed to equilibrate for an additional $5\times10^4$ steps. Production simulations are then performed for $1\times10^7$ steps, with particle coordinates and dipole moment vectors recorded every 100 steps. We determine the bulk liquid density from the resulting phase-separated configurations at different values of $\mu$ and $d$.

To further analyze the homogeneous liquid phase, we conduct bulk simulations at the coexistence liquid density determined from the slab calculations. The simulations are carried out in a cubic box  with dimensions adjusted to maintain the coexistence liquid density for 3000 uniformly distributed particles. At large dipole moments the liquid can reach very high densities, and simulations may diverge because of overlapping molecules, even following energy minimization. To mitigate this issue, we initialize the system with the box length in the $x$--direction set to 1.5 times the target value and gradually compress this dimension over $5\times10^4$ steps. This is followed by an additional $5\times10^4$ equilibration steps. Finally, production simulations are carried out for $1\times10^7$ steps, with particle coordinates and dipole moment vectors recorded every 100 steps.

\subsubsection{Analysis}

We analyze the simulation trajectories using the MDAnalysis toolkit \cite{michaud-agrawalMDAnalysisToolkitAnalysis2011} and MDCraft .\cite{ye_mdcraft_2024} MDCraft is employed to accommodate the features of the shifted Stockmayer model by treating dipole orientations in addition to particle positions through a custom trajectory reader. In this work, we focus primarily on the interparticle positional and orientational correlations that determine the structural and dielectric properties of the system. These correlations are conveniently characterized using three key angles defined by Langenbach and co-workers ,\cite{langenbach_co-oriented_2017,langenbachRelativePermittivityDipolar2020} as illustrated in Fig.~\ref{fig:orientation}. 

For any pair of particles $1$ and $2$, we define two angles $\varphi_1$ and $\varphi_2$ between the dipoles and the displacement vector connecting two real particle centers:
\begin{equation}
    \cos\varphi_1 \equiv \xi_1 
    = 
    \frac{\boldsymbol{\mu}_1\cdot\mathbf{r}_{12}}
    {|\boldsymbol{\mu}_1|\,|\mathbf{r}_{12}|}\,,
    \label{eq:xi_1}
\end{equation}
\begin{equation}
    \cos\varphi_2 \equiv \xi_2 
    = 
    \frac{\boldsymbol{\mu}_2\cdot\mathbf{r}_{12}}
    {|\boldsymbol{\mu}_2|\,|\mathbf{r}_{12}|}\,,
    \label{eq:xi_2}
\end{equation}
together with the angle between the two dipole vectors,
\begin{equation}
    \cos\Theta 
    =
    \frac{\boldsymbol{\mu}_1\cdot\boldsymbol{\mu}_2}
    {|\boldsymbol{\mu}_1|\,|\boldsymbol{\mu}_2|}\,.
\end{equation}
The corresponding torsional angle $\gamma_{12}$ is then defined as
\begin{equation}
    \cos\gamma_{12} 
    =
    \frac{\cos\Theta-\xi_1\xi_2}
    {\sqrt{1-\left(\xi_1\right)^2}
     \sqrt{1-\left(\xi_2\right)^2}}\,.
\end{equation}

From these quantities we construct the multivariate pair distribution function
$\rho(\xi_1,\xi_2,\gamma_{12},r_{12})$, averaged over all particle pairs. Integration over selected degrees of freedom yields marginal distributions that allow specific structural features to be extracted. For example, the radial distribution function (RDF) is obtained as:

\begin{equation}
    g(r_{12}) =\frac{1}{4\pi r_{12}^2}\int d\xi_1 \int d\xi_2 \int d\gamma_{12}\,\rho(\xi_1,\xi_2,\gamma_{12},r_{12})\,.
\end{equation}

Orientational correlations resolved with respect to the intermolecular separation are similarly defined as
\begin{equation}
    P(r_{12},\xi_1)=\frac{1}{4\pi r_{12}^2}\int d\xi_2 \int d\gamma_{12}\,\rho(\xi_1,\xi_2,\gamma_{12},r_{12})\,.
\end{equation}

For Stockmayer--type fluids it is also common to examine the angular distribution function (ADF) ,\cite{langenbach_co-oriented_2017,kohnsRelativePermittivityStockmayerType2020}
\begin{equation}
    g(\xi_1)=\int dr_{12} \int d\xi_2 \int d\gamma_{12}\,\rho(\xi_1,\xi_2,\gamma_{12},r_{12})\,.
\end{equation}

To isolate effects within the first solvation shell, the radial integrals are truncated at $r_{12} = 1.5\sigma$, which corresponds, on average, to the location of the first minimum in the RDF function. We note that this correlation function possesses a useful symmetry: while the ADF describes the angle between the dipole of particle 1 and the displacement vector between a pair, the corresponding distribution for particle 2 satisfies $g(\xi_2)=g(-\xi_1)$ (or $g(\varphi_2)=g(\pi-\varphi)$). That is, it is simply the reflection of the first particle's ADF about $\xi=0$ ($\varphi=\frac{\pi}{2}$). Without loss of generality, we therefore consider $g(\xi)\equiv g(\xi_1)$ ($g(\varphi)\equiv g(\varphi_1)$) hereafter. By a similar token, we also omit the indices on the center-of-mass separation distance $r_{12}$.

\subsection{COFFEE Theory}
\label{sec:coffee}
To gain deeper insight into the shifted Stockmayer fluid and to provide an analytical description of its equilibrium properties, we apply the COFFEE (Co--Oriented Fluid Functional Equation for Electrostatic interactions) theory developed by Langenbach and co-workers .\cite{langenbach_co-oriented_2017,kohnsRelativePermittivityStockmayerType2020} A full derivation is provided in the original work; here we summarize the equations relevant to our implementation .\cite{walkerClapeyronjlExtensibleOpenSource2022} 

The COFFEE framework expresses the residual Helmholtz free energy $F_\mathrm{res}$ of a system with Lennard-Jones and shifted dipolar interactions as the sum of three contributions:
\begin{equation}
    \frac{F_\mathrm{res}}{NkT} = \frac{F_\mathrm{FF}}{NkT} + \frac{F_\mathrm{NF}}{NkT} + \frac{F_\mathrm{ref}}{NkT}\,,
\end{equation}
where $k$ is Boltzmann’s constant and the subscripts FF, NF, and ref denote the far-field, near-field, and reference contributions, respectively. 

In its original formulation, COFFEE theory employed the full Lennard--Jones fluid as the reference system .\cite{lafitteAccurateStatisticalAssociating2013} 
Because our simulations use a truncated Lennard-Jones potential with cutoff $r_c=2.5\sigma$, we instead adopt a reference free energy based on the perturbed truncated and shifted (PeTS) equation of state .\cite{heierEquationStateLennardJones2018} 

The dipolar free energy is separated into near- and far-field contributions under the assumption that shifting the dipole primarily affects interactions within a radius of $1.5\sigma$ .\cite{langenbach_co-oriented_2017} 
The far-field term follows the functional form proposed by Vrabec and Gross ,\cite{grossEquationofstateContributionPolar2006} employing a Pad\'e approximation \cite{stellThermodynamicPerturbationTheory1974}:
\begin{equation}
    F_\mathrm{FF} = \frac{F_2}{1-\frac{F_3}{F_2}}\,,
\end{equation}
where $F_2$ and $F_3$ are parameterized from simulation data. The near-field contribution explicitly accounts for the local orientational environment of a reference particle through the orientational distribution function:
\begin{figure*}[htpb]
    \centering
    \includegraphics[width=0.95\textwidth]{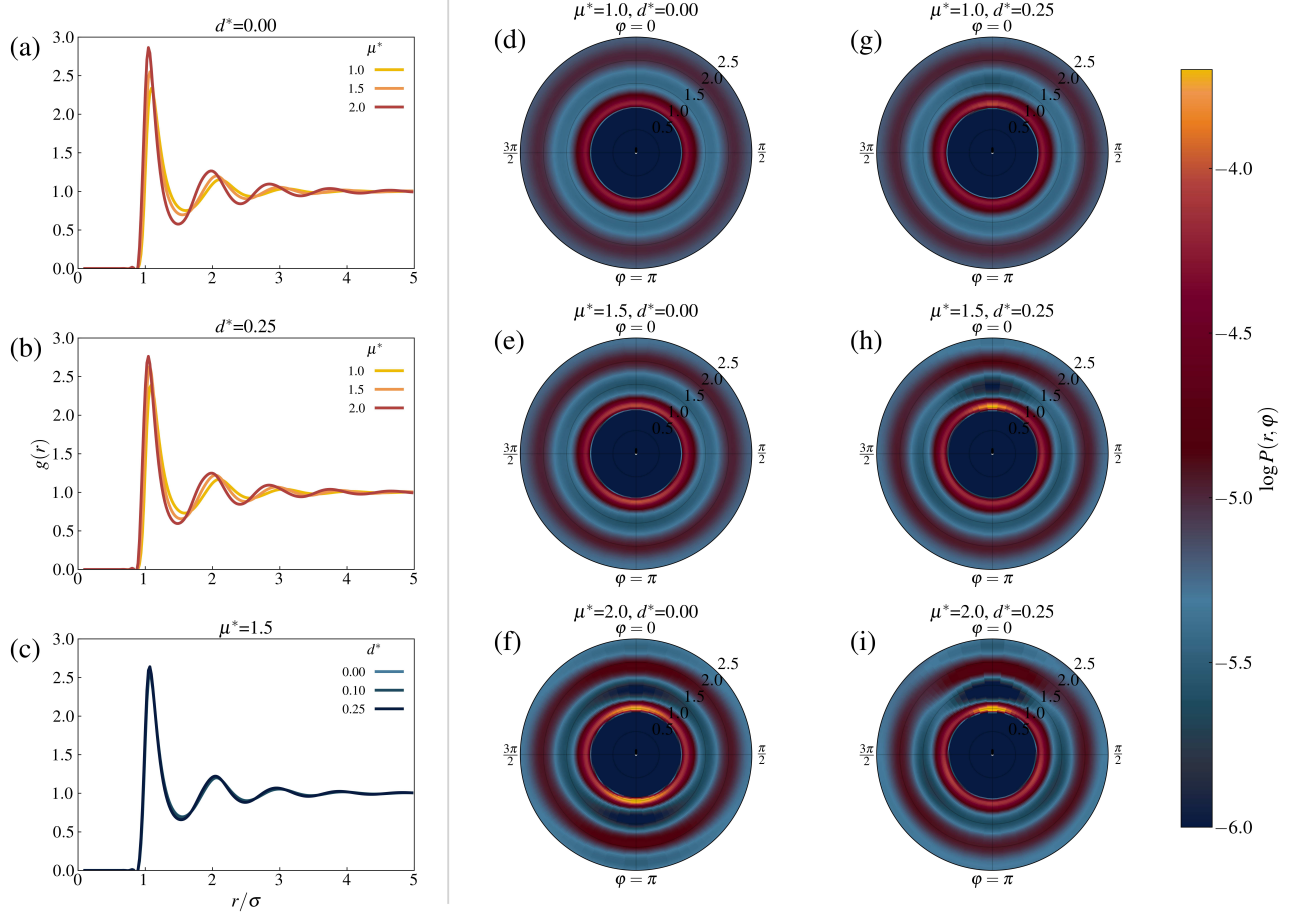}
    \caption{Radial distribution functions $g(r)$ in panels (a)–(c) and angular-radial distribution functions $\log P(r,\varphi)$ in panels (d)–(i) for the shifted Stockmayer fluid. In (a) and (b), $d^*=0.00$ and $d^*=0.25$, respectively, with varying dipole strengths; panel (c) fixes $\mu^*=1.500$ and varies $d^*$. Angular–radial distribution functions $\log P(r,\varphi)$ in polar coordinates. Color denotes the log-normalized probability of finding a neighbor LJ center at separation $r/\sigma$ and angle $\varphi$ between a reference particle dipole and its neighbor.}
    \label{fig:rdf-circ}
\end{figure*}
\begin{equation}
\label{eq:odf}
    O(\xi_1,\xi_2,\gamma_{12}) = \frac{1}{Q}\exp{\left(-\frac{24}{19}I_{\mu\mu}^\mathrm{NF}\int dr_{12}\beta\phi_\mathrm{dipole}(r_{12},\xi_1,\xi_2,\gamma_{12})\right)}\,.
\end{equation}

In Eq.~\eqref{eq:odf}, the normalization constant $Q$, interpreted as an orientational partition function, is defined as
\begin{align}
\label{eq:part_fun}
    Q = &\int d\xi_1d\xi_2d\gamma_{12} \\\nonumber
   &\exp \bigg(-\frac{24}{19}I_{\mu\mu}^\mathrm{NF}(y^*)\int dr_{12}\beta\phi_\mathrm{dipole}(r_{12},\xi_1,\xi_2,\gamma_{12})\bigg)\,,
\end{align}
where the correction factor $I_{\mu\mu}^{\mathrm{NF}}(y^*)$ depends only on the reduced dipole strength $y^*$ \cite{marxPhaseEquilibriaMixtures2022} and is given by:
\begin{equation}
    I_{\mu\mu}^\mathrm{NF}(y^*)=a_1 \exp\left(a_2y^*+a_3y^{*2}\right)\,,
\end{equation}
with:
\begin{equation}
    y^* = \frac{4\pi\beta\mu^2\rho}{9}\,.
\end{equation}
Note that all radial integrals within the preceding equations are bound between $r=0$ and $r=1.5\sigma$, only considering the first solvation shell. Because a wider range of dipole magnitudes $\mu$ and shifts $d$ are explored here, and a different reference system is employed, the constants $a_n$ are refitted using angular distribution functions obtained from simulation.

From $Q$, the near-field free-energy contribution can be written as\footnote{Although this expression appears simpler than those originally presented by Langenbach and co--workers, it is analytically equivalent under the same assumptions. A full derivation is provided in the Supplementary Information.}:
\begin{equation}
    \frac{F_\mathrm{NF}}{NkT} = -\frac{19\pi}{12}\rho\sigma^3g_\mathrm{HS}(\sigma)\ln\frac{Q}{\Omega}\,,
\end{equation}
where $g_\mathrm{HS}(\sigma)$ is the hard--sphere radial distribution function at contact .\cite{boublikHardSphereEquationState1970} Here $\Omega$ denotes the total orientational phase--space volume; for uniformly distributed orientations on the unit sphere, $\Omega=4\pi$.

The resulting theory, which we refer to as COFFEE--PeTS\footnote{The authors do not condone or recommend feeding coffee to one's pets.}, is implemented in the open--source package Clapeyron.jl .\cite{walkerClapeyronjlExtensibleOpenSource2022} 
All fitted parameters are reported in the Supplementary Information.

\section{Results and Discussion}
\begin{figure*}[ht!]
    \centering
    \includegraphics[width=0.8\textwidth]{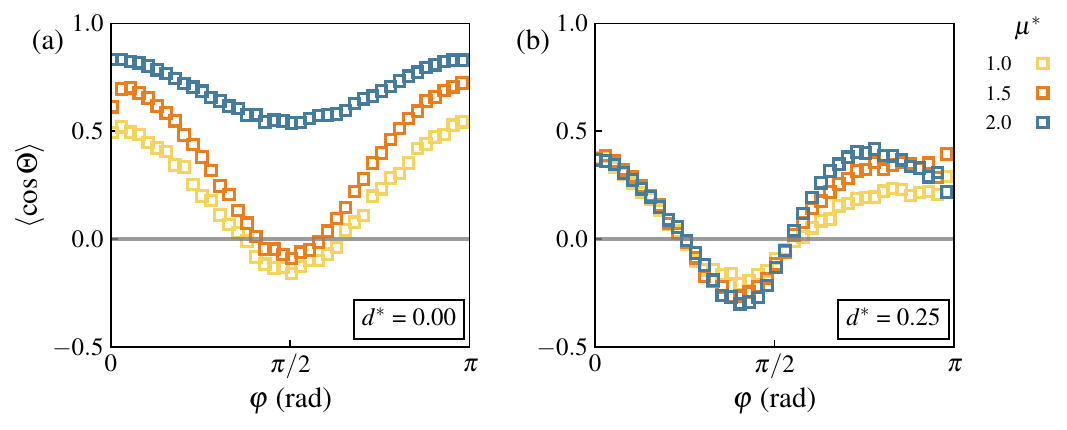}
    \caption{Average angle between dipoles of neighboring particles, $\langle\cos\Theta\rangle$, in the first solvation shell as a function of angular position $\varphi$ between a reference particle dipole and its neighboring particles for (a) $d=0.001$ with $\mu^*=1.000,\,1.500,\,2.000$ and (b) $d=0.250$ with the same $\mu^*$.} 
    \label{fig:or-rad}
\end{figure*}
We now examine the structural properties of the shifted Stockmayer fluid and the resulting dielectric behavior of homogeneous bulk phases. On the basis of these results, we subsequently analyze the vapor--liquid equilibrium and assess predictions obtained from the modified COFFEE equation.

Throughout this section, reduced Lennard--Jones units are employed. Accordingly, we define the dimensionless density, temperature, dipole moment, and shift as
\begin{equation}
    \rho^*=\rho\sigma^3,\quad
    T^* = \frac{kT}{\epsilon},\quad
    \mu^* = \frac{\mu}{\sqrt{4\pi\epsilon_0\epsilon\sigma^3}},\quad
    d^*=\frac{d}{\sigma},
\end{equation}
where the superscript $*$ denotes reduced quantities. As a reference, if $\epsilon/k=300\,\mathrm{K}$ and $\sigma=0.275\,\mathrm{nm}$, then $\mu^*\approx1.95$ and $\rho^*=0.9$ correspond approximately to liquid water near room temperature at $T^*=1$.

\subsection{Structural Properties}
\label{sec:struct}
A central objective of this study is to elucidate how shifting the dipole modifies the structure of the Stockmayer fluid. We therefore analyze several pair correlation functions: (1) the radial distribution function $g(r)$, (2) the angular-radial distribution $P(r,\varphi)$, (3) the angular dependence of dipole alignment $\langle \cos \Theta \rangle$ within the first solvation shell, and (4) the angular distribution function $g(\varphi)$. Note that all results obtained here pertain to liquid state of the sSF at vapor--liquid equilibrium conditions.

\begin{figure}[ht!]
    \centering
    \includegraphics[width=\linewidth]{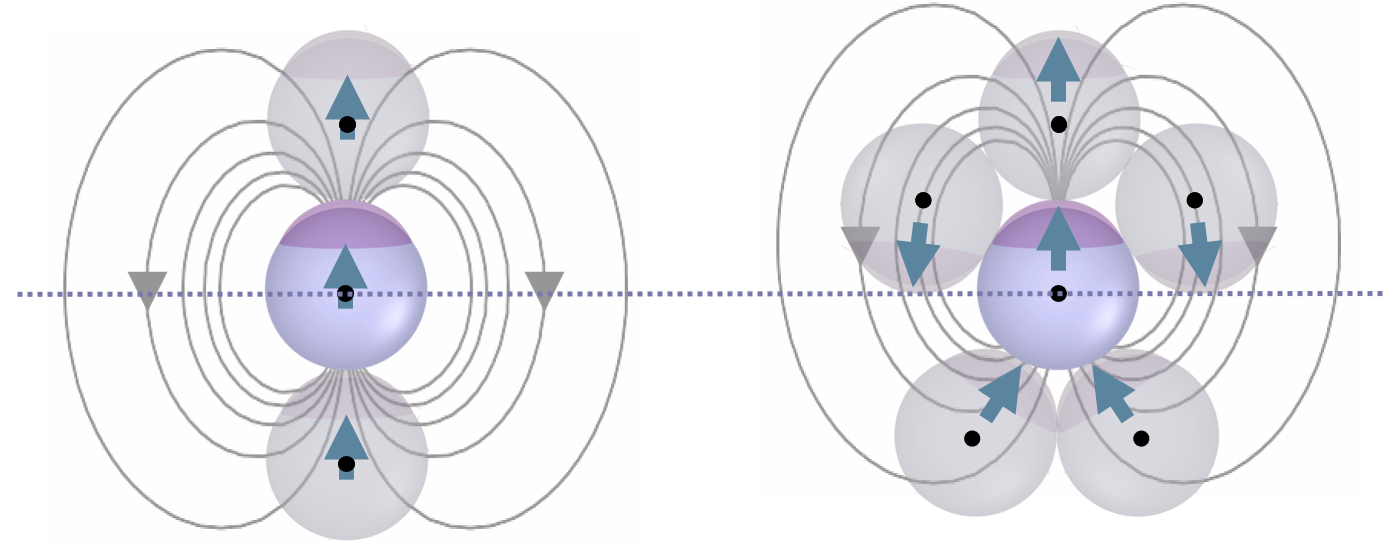}
    \caption{Schematic of the dipole field around a tagged particle for a non-shifted dipole (left) and a shifted dipole (right); the dotted midline is a visual guide to emphasize the shift. Black dots mark neighbor positions, with neighboring dipoles aligning with the local field. }
    \label{fig:schematic}
\end{figure}

The radial distribution function $g(r)$ describes the relative probability of finding a particle at a distance $r$ from a reference particle. For the shifted Stockmayer fluid, $g(r)$ reflects the combined influence of excluded--volume repulsion and dipolar forces. By comparing $g(r)$ for different dipole magnitudes and shifts, we assess how breaking the particle symmetry modifies short--range order. For an unshifted Stockmayer fluid, the radial structure is expected to be dominated by repulsive interactions ,\cite{shock_solvation_2020} such that $g(r)$ exhibits pronounced peaks near integer values of $r/\sigma$. This behavior is confirmed in Fig.~\ref{fig:rdf-circ}(a), where sharp oscillations indicate well-defined solvation shells. Increasing $\mu^*$ enhances local ordering, as reflected by the sharpening of the first peak and amplified subsequent oscillations. At intermediate dipole shifts, the radial distribution remains well defined, as shown in Fig.~\ref{fig:rdf-circ}(b) for $d^*=0.25$. As in the unshifted case, excluded-volume effects largely govern the radial ordering. This is further emphasized by Fig.~\ref{fig:rdf-circ}(c), which demonstrates that varying the dipole shift produces only minor changes in $g(r)$.

\begin{figure*}[ht!]
    \includegraphics[width=0.8\textwidth]{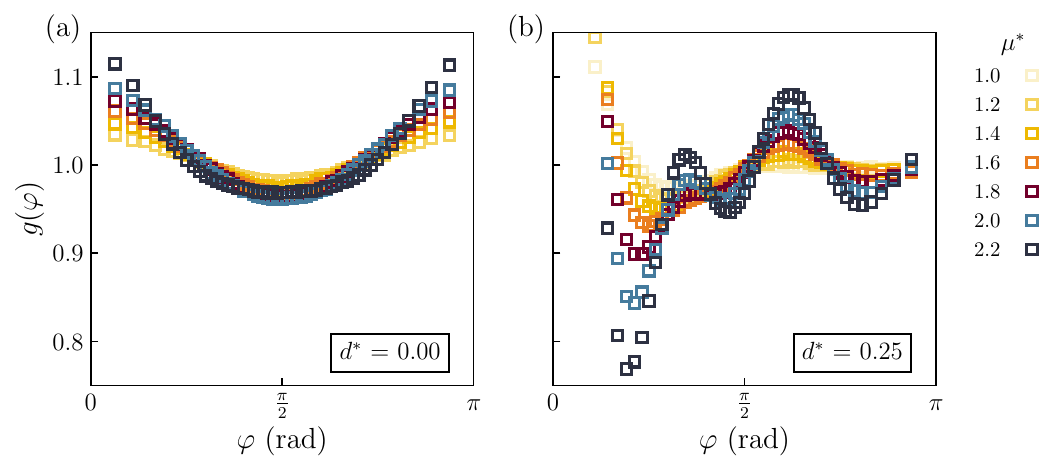}
    \caption{Angular distribution function $g(\varphi)$ obtained from simulations for varying dipole strength at constant shift (a) $d^*=0.00$ and (b) $d^*=0.25$. All systems were simulated at $T^*=1.0$ and at the co-existence liquid density.}
    \label{fig:adf}
\end{figure*}

Although the dipole shift does not substantially alter radial ordering, it introduces pronounced angular anisotropy in the solvation shells owing to the directional nature of the dipolar field. For small $\mu^*$ and minimal shift, excluded-volume effects dominate and the shells remain nearly spherical, consistent with the isotropic pattern in Fig.~\ref{fig:rdf-circ}(d). 
As $\mu^*$ increases, neighboring particles preferentially occupy regions near $\varphi=0$ and $\varphi=\pi$, where dipolar interactions are strongest, leading to the angular modulation observed in Fig.~\ref{fig:rdf-circ}(e) and (f).

To relate these positional preferences to orientational correlations, Fig.~\ref{fig:or-rad} shows the variation of $\langle \cos \Theta \rangle$ with angular position within the first solvation shell. For unshifted dipoles (Fig.~\ref{fig:or-rad}(a)), the distribution is symmetric about $\varphi=\pi/2$, reflecting the fore--aft symmetry of the dipolar field. Maxima in $\langle \cos \Theta \rangle$ appear near $\varphi=0$ and $\varphi=\pi$, where the field aligns with the reference dipole and strengthens with increasing $\mu^*$. Near $\varphi=\pi/2$, the field is weaker and favors anti-parallel alignment, producing a minimum in $\langle \cos \Theta \rangle$. At sufficiently large $\mu^*$ this minimum becomes positive, indicating global ferroelectric ordering in the liquid.

Introducing a dipole shift breaks this symmetry, as illustrated in Fig.~\ref{fig:rdf-circ}(g)--(i). The field near $\varphi=0$ becomes stronger than near $\varphi=\pi$, producing preferential accumulation at the head of the particle and broader angular distributions at the tail. These positional changes manifest in Fig.~\ref{fig:or-rad}(b): despite higher local density near $\varphi=0$, the average alignment is reduced, revealing orientational frustration induced by the shift. The effect is even more pronounced near $\varphi=\pi$

At larger $\mu^*$, an unexpected peak develops near $\varphi=3\pi/4$ in Fig.~\ref{fig:or-rad}(b). As illustrated schematically in Fig.~\ref{fig:schematic}, shifting the dipole displaces the region of strongest electric field away from $\varphi=\pi$. Neighboring particles therefore accumulate slightly off axis, where the field is strongest, broadening the distribution around $\varphi=\pi$ and producing a secondary maximum which becomes more pronounced at stronger dipole strengths.

\begin{figure}[ht!]
    \includegraphics[width=0.8\linewidth]{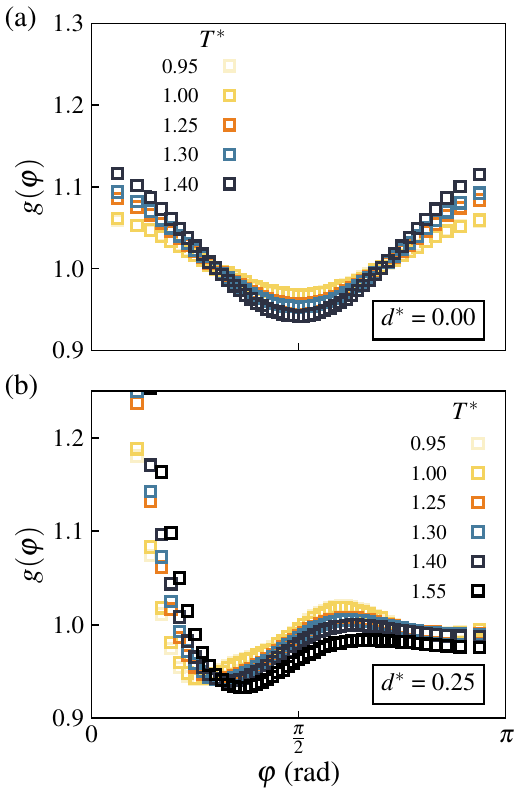}
    \caption{Angular distribution function obtained from simulations for systems with varying temperature and dipole shift. All systems were simulated with $\mu^*=1.5$ and at the co-existence liquid density.}
    \label{fig:adf_T}
\end{figure}

These trends are reflected in the angular distribution functions $g(\varphi)$ shown in Fig.~\ref{fig:adf}. Without a shift (Fig.~\ref{fig:adf}(a)), the ADF is symmetric about $\varphi=\pi/2$, consistent with Figs.~\ref{fig:or-rad} and \ref{fig:rdf-circ}. Increasing $\mu^*$ enhances the probability of finding neighbors near $\varphi=0$ and $\varphi=\pi$. Introducing a finite shift (Fig.~\ref{fig:adf}(b)) breaks this symmetry, first amplifying $g(\varphi)$ near $\varphi=0$ and subsequently producing a secondary maximum near $\varphi=3\pi/4$. A weaker peak near $\varphi=\pi/4$ also emerges at large $\mu^*$, consistent with the flattened angular distributions in Fig.~\ref{fig:rdf-circ}(g)--(i) and the reduced alignment seen in Fig.~\ref{fig:or-rad}. Following from Fig.~\ref{fig:or-rad}, we can see a minimum in $\langle\cos\Theta\rangle$ around these orientations, implying that these dipoles have a preference to be oriented anti-aligned with the reference particle. This is consistent with Fig.~\ref{fig:schematic} as, in this region, the dipolar field is aligned such that the preferred orientation would be anti-aligned with the reference dipole. 

Overall, introducing a dipole shift disrupts the preferential axial ordering characteristic of the conventional Stockmayer fluid at high dipole strengths, most clearly illustrated by the difference between Figs.~\ref{fig:or-rad}(a) and (b) at $\mu^*=2.0$, where $\langle\cos \Theta \rangle$ was previously $>0$ for all $\varphi$, with a shift, values $<0$ persist.

Finally, Fig.~\ref{fig:adf_T} illustrates the temperature dependence of the ADF at fixed $\mu^*=1.5$. Because these data correspond to coexistence liquid densities, increasing temperature is accompanied by a decrease in density. The reduced packing facilitates accumulation near energetically favorable orientations at $\varphi=0$ and $\varphi=\pi$, consistent with previous work .\cite{langenbach_co-oriented_2017} For shifted Stockmayer fluids, the ADF flattens, likely reflecting diminished steric constraints at lower density.

\begin{figure}[ht!]
    \centering
    \includegraphics[width=\linewidth]{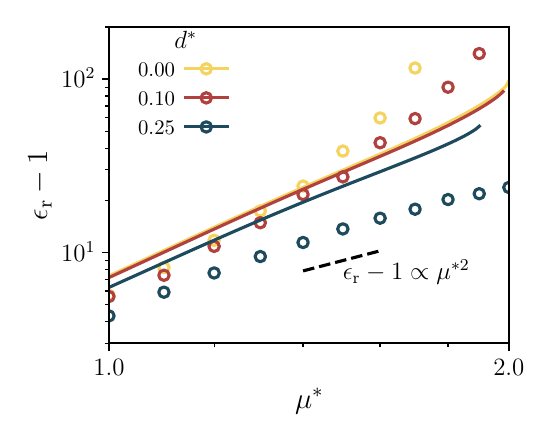}
    \caption{Dielectric constant of the varying dipole strength ($\mu^*$) and shift ($d^*$) at $ T^*$ = 1 and liquid density corresponding to vapor--liquid equilibrium conditions. Solid lines represent predictions made by COFFEE-PeTS while symbols were obtained from molecular dynamics simulations.}
    \label{fig:dielectric}
\end{figure}

\subsection{Dielectric Properties}
\label{sec:dielectric}
A key material property that depends strongly on dipolar correlations is the relative permittivity (dielectric constant), $\epsilon_\mathrm{r}$. The dielectric constant provides a macroscopic measure of the collective response of the fluid to an applied electric field and therefore reflects the extent of orientational correlations in the dipolar liquid. In this section we examine how $\epsilon_\mathrm{r}$ depends on the dipole strength $\mu^*$ and shift $d^*$. 
We compute $\epsilon_\mathrm{r}$ from equilibrium polarization fluctuations using the Kirkwood fluctuation formula ,\cite{neumann_dipole_1983}
\begin{equation}
    \epsilon_\mathrm{r} = 1+\frac{4\pi\rho}{3kT}(\langle\mathbf{M}^2\rangle-\langle\mathbf{M}\rangle^2)\,,
    \label{eq:Kirkwood}
\end{equation}
where the total system dipole moment is:
\begin{equation}
    \mathbf{M} = \sum_i\boldsymbol{\mu}_i\,.
\end{equation}

\begin{figure*}[ht!]
    \includegraphics[width=\textwidth]{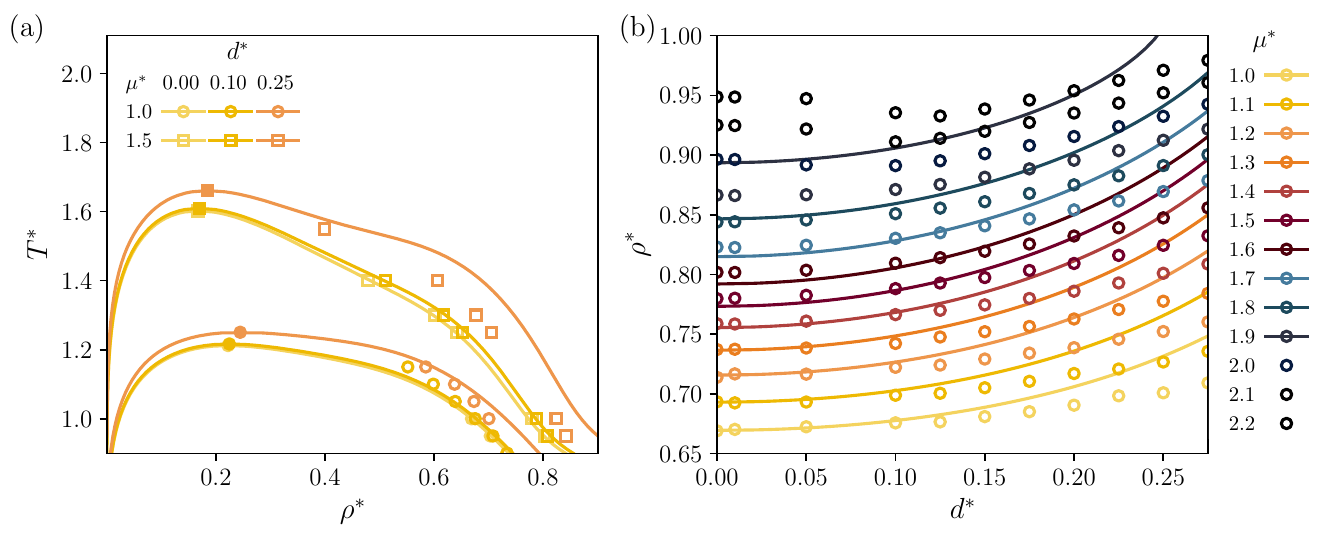}
    \caption{Liquid density at co-existence with varying dipole strength, shift and temperature. Figure (a) examines the full vapor--liquid envelope with increasing temperature, while Figure (b) considers the liquid density branch only at $T^*=1.0$ and varying $\mu^*$ and $d^*$. Solid lines represent predictions made by COFFEE-PeTS while symbols were obtained from molecular dynamics simulations.}
    \label{fig:vle}
\end{figure*}

For an isotropic, disordered fluid one expects $\langle \mathbf{M} \rangle \approx \mathbf{0}$.

Figure~\ref{fig:dielectric} reports $\epsilon_\mathrm{r}$ obtained from simulations at $T^*=1$ and at the coexistence liquid densities. For all cases, $\epsilon_\mathrm{r}$ increases with the dipole strength $\mu^*$, reflecting enhanced dipole--dipole correlations and stronger orientational ordering as $\mu^*$ increases. At the largest dipole moments, $\epsilon_\mathrm{r}$ rises sharply. This could be attributed to a ferroelectric transition. However, the system size is too small to fully resolve this transition.

Due to the shifting the dipole increasing the local interaction asymmetry, introducing a finite shift $d^*$ disrupts the cooperative dipole alignment and leads to a systematic reduction in $\epsilon_\mathrm{r}$. In the limit of large shifts, $\epsilon_\mathrm{r}$ approaches the mean--field result for noninteracting dipoles (the Debye--Langevin limit), shown as a dashed line in Fig.~\ref{fig:dielectric}. This suggests that, for sufficiently large $d^*$, local dipole rearrangements do not build a strong self--consistent reaction field and the dielectric response is dominated by the single--particle orientational susceptibility rather than collective correlations. This suppression is consistent with the structural trends discussed in the previous section: increasing $d^*$ reduces the average alignment between neighboring dipoles (Fig.~\ref{fig:or-rad}) and broadens the angular distributions (Fig.~\ref{fig:adf}), as illustrated schematically in Fig.~\ref{fig:schematic}.

Using the reparameterized COFFEE--PeTS model, we also predict $\epsilon_\mathrm{r}$ at these state points. Although quantitative agreement deteriorates at large $\mu^*$ and $d^*$ (conditions beyond the regime for which COFFEE was originally developed), the theory reproduces the qualitative trends and highlights the direct connection between local orientational structure and the macroscopic dielectric response.

\subsection{Vapour--Liquid Equilibria}
\label{sec:vle}
Linking microscopic structure to macroscopic behavior, we examine the coexistence liquid densities obtained from simulation and compare them with predictions from COFFEE--PeTS in Fig.~\ref{fig:vle}. We restrict application of COFFEE--PeTS to intermediate values of the dipole strength and shift, since larger values lie beyond the regime for which the original COFFEE theory was developed.

Considering first the temperature dependence in Fig.~\ref{fig:vle}(a), the coexistence liquid density increases systematically with dipole strength at a given temperature. Stronger dipoles enhance cohesive interactions, promoting closer packing and raising the critical temperature. These trends are consistent with previous studies \cite{grossEquationofstateContributionPolar2006} and are reproduced well by COFFEE--PeTS within its intended range of applicability. The introduction of a dipole shift, through competition between enhanced local interactions and the orientational frustration, results in a modest increase in the co-existing liquid density. 
This balance is also captured qualitatively by COFFEE--PeTS. 
At larger shifts, however, the quantitative performance of the theory deteriorates, as is particularly evident in Fig.~\ref{fig:vle}(b). In this regime, COFFEE--PeTS reproduces coexistence densities with high fidelity only for small $d^*$ and moderate $\mu^*$.

At very large dipole strengths and small displacements, Fig.~\ref{fig:vle}(b) shows that increasing $d^*$ initially leads to a decrease in the liquid density. At such high $\mu^*$ the unshifted Stockmayer fluid develops strong global polarization and highly ordered local packing, which results in unusually dense liquid phases, consistent with Fig.~\ref{fig:or-rad} and previous studies. Introducing even a small displacement disrupts this ordering, suppressing the formation of the polarized state and thereby reducing the liquid density. 

Considering Figs.~\ref{fig:dielectric} and \ref{fig:vle} together reveals an intriguing result. Although introducing a dipole shift effectively strengthens cohesive interactions and increases the liquid density, the dipolar correlations, and consequently the dielectric constant, are suppressed. This behavior is counterintuitive: according to the Kirkwood fluctuation formula, stronger intermolecular interactions and denser packing would typically enhance collective dipole fluctuations and thus increase the dielectric constant. Using the physical intuition developed in the previous section, we can rationalize this behavior: while the dipole shift increases the magnitude of local interactions, it simultaneously introduces orientational frustration within the first solvation shell, thereby weakening cooperative dipole alignment and suppressing long-range polarization.

\section{Conclusion}
In this work, we have combined molecular simulations with analytical theory to demonstrate that even a modest shift of the dipole off the center of a Stockmayer particle can profoundly alter its microscopic structure, dielectric response, and macroscopic phase behavior. Although aspects of shifted dipoles have been explored previously, our results provide a unified physical picture that connects local symmetry breaking in pair correlations to emergent thermodynamic consequences.

The central effect of shifting the dipole is the loss of fore-aft symmetry in the local environment surrounding each particle. While radial distribution functions remain largely governed by excluded-volume packing, the angular structure of the solvation shell changes dramatically: the dipolar field is strengthened at the head of the particle ($\varphi=0$) and weakened at the tail ($\varphi=\pi$), producing preferential accumulation and alignment in the forward direction and frustrated orientational correlations in the rear. At sufficiently strong dipoles, this frustration becomes particularly pronounced, shifting preferred configurations away from axial positions and disrupting the highly ordered structural characteristic of the conventional Stockmayer fluid.

These microscopic asymmetries propagate directly to collective electrostatic behavior. Despite stronger local interactions, introducing a dipole shift weakens cooperative dipolar correlations, leading to a systematic reduction in the dielectric constant. For large shifts, the dielectric response approaches the mean--field Debye limit, indicating that local dipole rearrangements no longer generate a strong self-consistent reaction field and that collective polarization is effectively suppressed, reducing the dielectric constant by an order of magnitude.

The consequences of dipolar asymmetry extend to phase behavior. Increasing dipole strength enhances cohesion and raises the critical temperature, as in the standard Stockmayer model, but the dipole shift fundamentally reshapes this trend. At intermediate dipole strengths, the introduction of a shift further enhances cohesion and raises the liquid density. However, at high dipole strengths, where the unshifted fluid develops strongly polarized, densely packed liquid phases, the introduction of even a small shift frustrates this ordering, suppresses the formation of a ferroelectric--like state, and lowers the liquid density. Conversely, for disordered fluid phases, the increased attraction between shifted dipoles increases the liquid density. In this way, the dipole shift acts not simply as a quantitative perturbation but as a qualitative control parameter governing structure, dielectric response, and coexistence.

Taken together, our results reveal a striking principle: modifying the \textit{geometry} of electrostatic interactions can be more disruptive to the liquid structure than the raw magnitude of the interactions themselves. This sensitivity to dipole placement opens new avenues for designing model fluids and coarse-grained representations of polar materials in which dielectric properties, phase behavior, and interfacial structure can be tuned independently. Such control is likely to be relevant for a broad class of systems, including polar solvents, electrolytes, ionic liquids, and soft interfaces, where displaced charges and anisotropic electrostatic environments are ubiquitous. Future work will extend these ideas to multicomponent mixtures and confined geometries, where geometric frustration of electrostatics may play an even more dramatic role.
\begin{acknowledgments}
The authors would like to thank Dr. Benjamin Ye for helping develop much of the tools required for the analysis used in this work. S.V. is supported by the U.S. Department of Energy, Office of Science, Office of Advanced Scientific Computing Research, Department of Energy Computational Science Graduate Fellowship under Award Number DE-SC0022158. Partial support for this research is provided by Hong Kong Quantum AI Lab, AIR@InnoHK of Hong Kong Government. B.Z. gratefully acknowledges support by Harvey Mudd College start-up grant and NSF CAREER Award CHE-2337602. A.V. acknowledges the Resnick Sustainability Institute and the WAVE Fellows Program at Caltech for their support.

\end{acknowledgments}

\section*{Data Availability Statement}
The data that support the findings of this study are available from the corresponding author upon reasonable request.

\section*{References}
\bibliography{references,stockmayer_bulk,extra}

\end{document}


\preprint{AIP/123-QED}

\def\*#1{\mathbf{#1}}
\def\&#1{\mathrm{#1}}

\title{Supplementary Information for\textit{ Stockmayer Fluid with a Shifted Dipole: Bulk Behavior}}

\author{Pierre J. Walker}
\altaffiliation{These authors contributed equally.}
\affiliation{Division of Chemistry and Chemical Engineering, California Institute of Technology, Pasadena, CA 91125, United States}

\author{Ananya Venkatachalam}
\altaffiliation{These authors contributed equally.}
\affiliation{Department of Chemistry, Harvey Mudd College, Claremont, CA 91711, United States}

\author{Samuel Varner}
\altaffiliation{These authors contributed equally.}
\affiliation{Division of Chemistry and Chemical Engineering, California Institute of Technology, Pasadena, CA 91125, United States}

\author{Bilin Zhuang}
\affiliation{Department of Chemistry, Harvey Mudd College, Claremont, CA 91711, United States}
\email{bzhuang@g.hmc.edu}

\author{Zhen-Gang Wang}
\affiliation{Division of Chemistry and Chemical Engineering, California Institute of Technology, Pasadena, CA 91125, United States}
\email{zgw@caltech.edu}

\date{\today}

\begin{abstract}

\end{abstract}

\maketitle
\subsubsection*{Simplification of the COFFEE near-field free energy}
The orientational distribution function in COFFEE\cite{langenbach_co-oriented_2017} is given as:
\begin{equation}
    O(\xi_1,\xi_2,\gamma_{12}) = \frac{1}{Q}\exp{\left(-\frac{24}{19}I_{\mu\mu}^\mathrm{NF}\int dr_{12}\beta\phi_{\mu\mu}(r_{12},\xi_1,\xi_2,\gamma_{12})\right)}
\end{equation}
The original expression for the near-field free energy within COFFEE is written as:
\begin{align}
\frac{F_\mathrm{NF}}{NkT} &= \frac{19\pi}{12} \rho^* g_{\text{hs}}(\sigma) \int d\boldsymbol{\omega_1} \int d\boldsymbol{\omega_2} \, O(\boldsymbol{\omega_1}, \boldsymbol{\omega_2}) \ln\left( \Omega O(\boldsymbol{\omega_1}, \boldsymbol{\omega_2}) \right) \\\nonumber
& + 2\pi \rho^* \frac{{\mu^*}^2}{T^*} g_{\text{hs}}(\sigma) I_{\mu\mu} \int_1^{3/2} dr^* \int d\boldsymbol{\omega_1} \int d\boldsymbol{\omega_2} \, O(\boldsymbol{\omega_1}, \boldsymbol{\omega_2}) o(r^*, \boldsymbol{\omega_1}, \boldsymbol{\omega_2})
\end{align}
Factoring out the common multipliers:
\begin{align}
\frac{F_\mathrm{NF}}{NkT} &= 2\pi\rho^*\bigg[\frac{19}{6} g_{\text{hs}}(\sigma) \int d\boldsymbol{\omega_1} \int d\boldsymbol{\omega_2} \, O(\boldsymbol{\omega_1}, \boldsymbol{\omega_2}) \ln\left( \Omega O(\boldsymbol{\omega_1}, \boldsymbol{\omega_2}) \right) \\\nonumber
& + \frac{{\mu^*}^2}{T^*} g_{\text{hs}}(\sigma) I_{\mu\mu} \int_1^{3/2} dr^* \int d\boldsymbol{\omega_1} \int d\boldsymbol{\omega_2} \, O(\boldsymbol{\omega_1}, \boldsymbol{\omega_2}) o(r^*, \boldsymbol{\omega_1}, \boldsymbol{\omega_2})\bigg]
\end{align}
Substituting the expression for $O(\xi_1,\xi_2,\gamma_{12})$ and separating the terms which do not depend on $\xi_1$,$\xi_2$ and $\gamma_{12}$ (recalling that $\int d\boldsymbol{\omega_1} \int d\boldsymbol{\omega_2} \, O(\boldsymbol{\omega_1}, \boldsymbol{\omega_2}) = 1$):
\begin{align}
\frac{F_\mathrm{NF}}{NkT} &= 2\pi \rho^* g_{\text{hs}}(\sigma) \Bigg[ \frac{19}{24} \Bigg( \ln\left( \frac{\Omega}{Q} \right)  \\\nonumber & +\int d\boldsymbol{\omega_1} \int d\boldsymbol{\omega_2} \, O(\boldsymbol{\omega_1}, \boldsymbol{\omega_2})
 \ln \left( \exp\left(-\frac{24}{19} \frac{{\mu^*}^2}{T^*}  I_{\mu\mu} \int_1^{3/2} dr^* \, o(r^*, \boldsymbol{\omega_1}, \boldsymbol{\omega_2}) \right) \right)  \Bigg) \\
& +  \frac{{\mu^*}^2}{T^*} I_{\mu\mu} \int_1^{3/2} dr^* \int d\boldsymbol{\omega_1} \int d\boldsymbol{\omega_2} \, O(\boldsymbol{\omega_1}, \boldsymbol{\omega_2}) o(r^*, \boldsymbol{\omega_1}, \boldsymbol{\omega_2}) \Bigg]\nonumber
\end{align}
Expanding the logarithm:
\begin{align}
\frac{F_\mathrm{NF}}{NkT} &= 2\pi \rho^* g_{\text{hs}}(\sigma) \Bigg[ \frac{19}{24} \Bigg(\ln\left( \frac{\Omega}{Q} \right) 
 \\\nonumber
& - \frac{24}{19}\frac{{\mu^*}^2}{T^*} I_{\mu\mu} \int d\boldsymbol{\omega_1} \int d\boldsymbol{\omega_2} \,  \int_1^{3/2} dr^* \, O(\boldsymbol{\omega_1}, \boldsymbol{\omega_2}) o(r^*, \boldsymbol{\omega_1}, \boldsymbol{\omega_2})  \Bigg) \\
& + \frac{{\mu^*}^2}{T^*} I_{\mu\mu} \int_1^{3/2} dr^* \int d\boldsymbol{\omega_1} \int d\boldsymbol{\omega_2} \, O(\boldsymbol{\omega_1}, \boldsymbol{\omega_2}) o(r^*, \boldsymbol{\omega_1}, \boldsymbol{\omega_2}) \Bigg]\nonumber
\end{align}
It is now easy to see that second and third term perfectly cancel, giving:
\begin{equation}
    \frac{F_\mathrm{NF}}{NkT}  = -\frac{19\pi}{12}\rho\sigma^3g_\mathrm{HS}(\sigma)\ln\frac{Q}{\Omega}
\end{equation}


\subsection*{COFFEE-PeTS parameters}
Within our implementation of COFFEE-PeTS, three sets of empirical parameters are needed. The first is for the function $I_{\mu\mu}^\mathrm{NF}$:
\begin{equation}
    I_{\mu\mu}^\mathrm{NF} = a_0\exp(a_1y^*+a_2(y^*)^2)\,.
\end{equation}
The second is in the far-field term which requires a function $J_2$:
\begin{equation}
    J_2 = \sum_{n=0}^4\left(b_{2,n}+c_{2,n}\frac{1}{T^*}\right)\eta^n\,,
\end{equation}
where $T^*=kT/\epsilon$ and $\eta=\frac{\pi}{6}\rho\sigma^3$. The final set of parameters is for the function $J_3$:
\begin{equation}
    J_3 = \sum_{n=0}^4\left(b_{3,n}+c_{3,n}\frac{1}{T^*}\right)\eta^n\,.
\end{equation}
These parameters are given in Table \ref{tbl:params}.

\begin{table}[]
\caption{COFFEE-PeTS equation parameters.}
\label{tbl:params}
\begin{tabular}{|l|ccccc|}\hline
$n=$       & 0          & 1         & 2          & 3          & 4      \\\hline
$a_n$      & 1.0889     & -0.25908  & 0.0084889  & 0.00000    & 0.00000 \\
$b_{2,n}$  & 0.16458    & -0.1429   & 0.95093    & 0.31398    & -2.6070 \\
$c_{2,n}$  & 0.57161    & -2.3623   & 2.1000     & 0.00000    & 0.00000 \\
$b_{3,n}$  & -0.0019945 & -0.030557 & 0.017623   & -0.0010382 & 0.00000 \\
$c_{3,n}$  & -0.0047808 & 0.039684  & -0.0002983 & 0.00000    & 0.00000 \\\hline
\end{tabular}
\end{table}

\bibliography{references}